\documentclass[twocolumn,10pt]{aastex63}

\usepackage{color}

\received{December, 7, 2020}
\accepted{March 31, 2021}
\submitjournal{ApJ}
\shortauthors{S.Sato et al.}
\graphicspath{{./}{figures/}}

\begin{document}

\title{Machine learning application to Fermi-LAT data: sharpening all-sky map and emphasizing variable sources}

\correspondingauthor{Shogo Sato}
\email{s.shogo19961111@akane.waseda.jp}

\author{Shogo Sato}
\affiliation{Faculty of Science and Engineering, Waseda University, 3-4-1 Ohkubo, Shinjuku, Tokyo, 169-8555, Japan}

\author{Jun Kataoka}
\affiliation{Faculty of Science and Engineering, Waseda University, 3-4-1 Ohkubo, Shinjuku, Tokyo, 169-8555, Japan}

\author{Soichiro Ito}
\affiliation{Faculty of Science and Engineering, Waseda University, 3-4-1 Ohkubo, Shinjuku, Tokyo, 169-8555, Japan}

\author{Jun'ichi Kotoku}
\affiliation{Teikyo University, 2-11-1 Kaga, Itabashi, Tokyo, 173-8605, Japan}

\author{Masato Taki}
\affiliation{Rikkyo University, 3-34-1, Nishi-ikebukuro, Toshima, Tokyo, 171-8501, Japan}

\author{Asuka Oyama}
\affiliation{Teikyo University, 2-11-1 Kaga, Itabashi, Tokyo, 173-8605, Japan}

\author{Takaya Toyoda}
\affiliation{Faculty of Science and Engineering, Waseda University, 3-4-1 Ohkubo, Shinjuku, Tokyo, 169-8555, Japan}

\author{Yuki Nakamura}
\affiliation{Faculty of Science and Engineering, Waseda University, 3-4-1 Ohkubo, Shinjuku, Tokyo, 169-8555, Japan}

\author{Marino Yamamoto}
\affiliation{Faculty of Science and Engineering, Waseda University, 3-4-1 Ohkubo, Shinjuku, Tokyo, 169-8555, Japan}

\begin{abstract}
A novel application of machine-learning (ML) based image processing algorithms is proposed to analyze an all-sky map (ASM) obtained using the Fermi Gamma-ray Space Telescope. An attempt was made to simulate a one-year ASM from a short-exposure ASM generated from one-week observation by applying three  ML based image processing algorithms: dictionary learning, U-net, and Noise2Noise. Although the inference based on ML is less clear compared to standard likelihood analysis, the quality of the ASM was generally improved. In particular, the complicated diffuse emission associated with the galactic plane was successfully reproduced only from one-week observation data to mimic a ground truth (GT) generated from a one-year observation. Such ML algorithms can be implemented relatively easily to provide sharper images without various assumptions of emission models. In contrast, large deviations between simulated ML  maps and GT map were found, which are attributed to the significant temporal variability of blazar-type active galactic nuclei (AGNs) over a year. Thus, the proposed ML methods are viable not only to improve the image quality of an ASM, but also to detect variable sources, such as AGNs, algorithmically, i.e., without human bias. Moreover, we argue that this approach is widely applicable to ASMs obtained by various other missions; thus, it has the potential to examine giant structures and transient events, both of which are rarely found in pointing observations.
\end{abstract}


\section{Introduction} \label{sec:intro}
All-sky monitoring in astronomy has begun a new era in which continuous monitoring of persistent sources as well as automated searches to discover new transient is possible. Moreover, the all-sky monitor is a unique approach that reveals the giant structures that are rarely seen in pointing observations. In the 1990s, ROSAT provided the first precise all-X-ray sky map below 2 keV \citep{Snowden_1997}. The observation started in July 1990 and lasted for approximately six months. The RXTE-ASM \citep{Bradt_1993} and MAXI onboard the International Space Station \citep{Matsuoka_2009} carry one-dimensional slat/shadow cameras extending the maximum energy window to 10 keV, which also makes it possible to monitor the brightest X-ray sources on a daily or hourly basis. More recently, eROSITA on the Spectrum Roentgen Gamma \citep{Merloni_2012} observatory, launched in June 2019, has provided a new, sharp view of hot and energetic processes taking place across the universe. For gamma rays, the Fermi Gamma-ray Space Telescope was launched in 2008 \citep{Atwood_2009}. It can observe 30-MeV to 100-GeV  gamma-ray sky with unprecedented sensitivity. Owing to a large field of view of 2.4 str, the large-area telescopes (LATs) onboard Fermi can monitor the sky every 3-hrs.

Along with the continuous monitoring of persistent sources, various missions were planned for detecting transient sources, such as gamma-ray bursts (GRBs) and/or flaring activity of active galactic nuclei (AGNs). For example, BATSE onboard CGRO \citep{Fishman_1992} carried eight scintillation detectors to cover the entire sky, localizing more than 2700 GRBs in the 10-year  observation. HETE2 \citep{Kippen_2003} and Swift \citep{Gehrels_2004} are also missions dedicated to finding and providing quick localization of GRBs by means of prompt follow-up X-ray and optical/UV observations. In general, these all-sky observations provide a large amount of data; thus, it is not easy to pick up interesting events or phenomena without overlooking some. Moreover, when the exposures for specific fields are very short, detection of persistent sources and/or transient events may suffer from low photon statistics, especially when the background dominates the signal. Development of a new protocol dedicated to the analysis of all-sky monitoring is required to discriminate real and unreal events effectively and quickly without any human bias.

In this context, various analytical methods have been developed for source identification and modeling of foreground diffuse gamma-ray emission (DGE). In DGE modeling, the cosmic-ray (CR) intensities and spectra are modeled by considering their propagation in the galaxy \citep{Ackermann_2008, Ackermann_2012} . In contrast, Bayesian-inference-based analysis reconstructs flux contributions to the diffuse and point-like sources \citep{Guglielmetti_2009}. In addition, the D3PO-inference algorithm reaches de-noising, de-convoluting, and decomposing \citep{Selig_2015}. For variable source detection, the Fermi data are processed using automated science processing (ASP) in the LAT Instrument Science Operations Center (ISOC) \citep{Robert_2012}. ASP searches for flares by calculating the flux using maximum-likelihood fitting. Apparently, this is the most accurate approach when sources on the catalog are filtered to some extent; however, developing and installing such an algorithm is challenging for individual users. In this context, aperture photometry light curves \citep{Lenain_2017} and Fermi all-sky variability analysis (FAVA) \citep{Ackermann_2013, Abdollahi_2017} are simple and fast analytical methods for variable source search. These methods are convenient; however, the quality of the images is not improved.

 As for the imaging analysis, significant progress has been made in machine learning (ML), especially deep learning, to find feature values in images of interest, for application in face recognition systems. In medical applications, similar approaches have already been implemented to determine lesion regions — for example, based on computed tomography images \citep{Kalinovsky_2017}. An obvious advantage of using ML is that it reads images in a short time with high accuracy without any human bias in the most ideal cases. However, similar ML techniques have been implemented only very recently in astronomical research. For example, a new computer program, Morpheus, has been developed to analyze astronomical image data pixel-by-pixel to identify and classify all galaxies and stars \citep{Hausen_2020}. In addition, spectral modeling of a type II supernova based on the ML approach is being developed \citep{Vogl_2020}. Application of ML techniques to all-sky survey data obtained by X-ray and gamma-ray satellite missions is also crucial, not only to best allocate resources for the data analysis, but also to examine new astronomical phenomena that have yet to be explored.

In this study, we proposed a new ML approach (1) to improve the quality of low-statistic images and (2) find variable sources and possibly even transient sources automatically from all-sky gamma-ray maps obtained with Fermi-LAT. In particular, short  observation data (one-week integration) were prepared as input, and long- observation data (one-year integration) were predicted using three independent  ML algorithms — dictionary learning \citep{Kreutz_2003}, U-net \citep{Ronneberger_2015}, and Noise2Noise \citep{Lehtinen_2018} — for detailed/quantitative comparison with actual observational data. The remainder of this article is organized as follows. In Section~\ref{sec:2}, the details of the Fermi-LAT analysis and data processing before the input  ML algorithms are presented. In Section~\ref{sec:3}, the architectures of the three  ML algorithms are described. The results of ML predictions are presented in Section~\ref{sec:4}. In Section~\ref{sec:5}, all-sky maps (ASMs) generated by ML algorithms and ground-truth (GT) map are compared in detail in selective regions to examine the possible origin of the mismatch. In Section~\ref{sec:6}, the conclusions are presented. 

\section{OBSERVATIONS AND ANALYSIS} \label{sec:2}
\subsection{Fermi-LAT observations} \label{subsec:2-1}
In this study, ML  based algorithms were applied to a gamma-ray ASM observed by Fermi-LAT. The first objective was to fabricate a high-quality ASM with sufficient photon statistics such that the ASM uncertainties are small compared to the counting uncertainties of shorter time-scale maps when only a low-quality ASM is known and/or provided as an input. As discussed in section~\ref{sec:3} ,  dictionary learning and U-net require the image pairs of short-observation map and long-observation map for image prediction. Moreover, because the Noise2Noise demands image pairs with different noise patterns, two types of  short-observation maps, corresponding to the first and second week in each year, were prepared. Therefore, a long-term observation map and two types of short-term observation maps in the energy range from 100 MeV to 300 GeV were created at the beginning of each year for years 2009 to 2019’ In this study, all energy values were incorporated into a single map. We produced the ASMs using gamma rays in the \textsc{P8R3\_SOURCE} event selection and the associated \textsc{P8R3\_SOURCE\_V2} instrument response functions (IRFs). In addition, to correct the ASM for exposure, we calculated an exposure map with one energy bin using {\tt gtexpcube2}, which is the standard analysis tool provided with Fermi-LAT ScienceTools. We also divided the ASM by the generated exposure map. In addition, the zenith angle cut was set to 90$^{\circ}$ to remove gamma rays from the Earth’s limb in the analysis. In this study, ASMs were produced in the galactic coordinates in the Hammer-Aitoff projection. We generated count maps with a resolution of 0.1 deg$^{2}$ per pixel.

The  ML algorithms assume that all sources are persistent over one year, keeping their fluxes exactly comparable to what has been observed in the first week. Thus, a large deviation of the  ML maps and the GT map may suggest that variable sources are included in regions of interest (ROIs). In such cases, a light curve was generated by using binned analysis to investigate the flux variability for specific gamma-ray sources. For each source, a 15$^{\circ}$ $\times$ 15$^{\circ}$ ROI centered at its 4FGL catalog \citep{Abdollahi_2020} was chosen, and a single power-law model was applied to the data. When creating the light curve, the prefactor  and the index of the central source were free, while the parameters of the other sources within the ROI were fixed. As the background model, the isotropic and diffuse backgrounds were modeled using \textsc{ISO\_P8R3\_SOURCE\_V 2\_v1.txt} and \textsc{gll\_iem\_v07.fits}, respectively. Moreover, only the flux bins with test static (TS) $>$ 9 (corresponding to $\sim$ 3$\sigma$) were used for each light curve.

\subsection{Preprocessing for ML}
The following four processing steps were applied to each ASM obtained in Section~\ref{subsec:2-1} before training the  ML algorithms:
\begin{enumerate}
    \item Applying Gaussian blurring
    \item Applying the natural logarithm function
    \item Standardization of the pixel values in each map
    \item Data augmentation
\end{enumerate}
First, Gaussian blurring was performed for each ASM to reduce the number of pixels whose value equals zero. The blurring process is necessary to prevent  ML algorithms from learning only  map smoothing. Subsequently, the natural logarithm function was applied to these maps owing to the high variance of pixel values; then, the pixel values were normalized from 0 to 1. Finally, random sampling was performed for data augmentation. In this process, the same regions were extracted from the ASM pairs with  short-observation and long-observation maps, and the extracted images were randomly flipped. The size of the extracted images was 64 $\times$ 64 pixels, which corresponds to $6^{\circ} \times 6^{\circ}$ in the galactic center. The random sampling was repeated 2000 times for each ASM pair. Detailed information about the components of the datasets is presented in Table~\ref{table1}. 
\begin{figure}
\centering
\includegraphics[width = 1.0\linewidth]{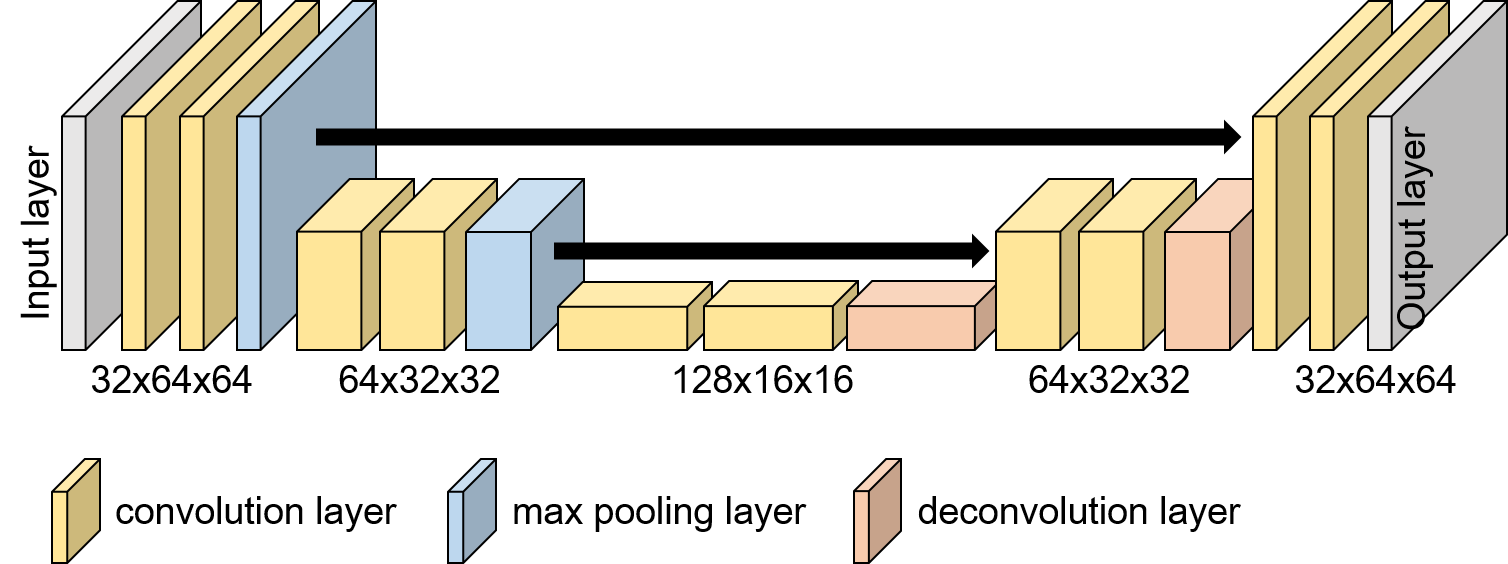}
\caption{\label{fig:unet} U-Net architecture with two downsampling layers: black arrows denote shortcut path}
\end{figure}
\begin{figure}
\centering
\includegraphics[width = 1.0\linewidth]{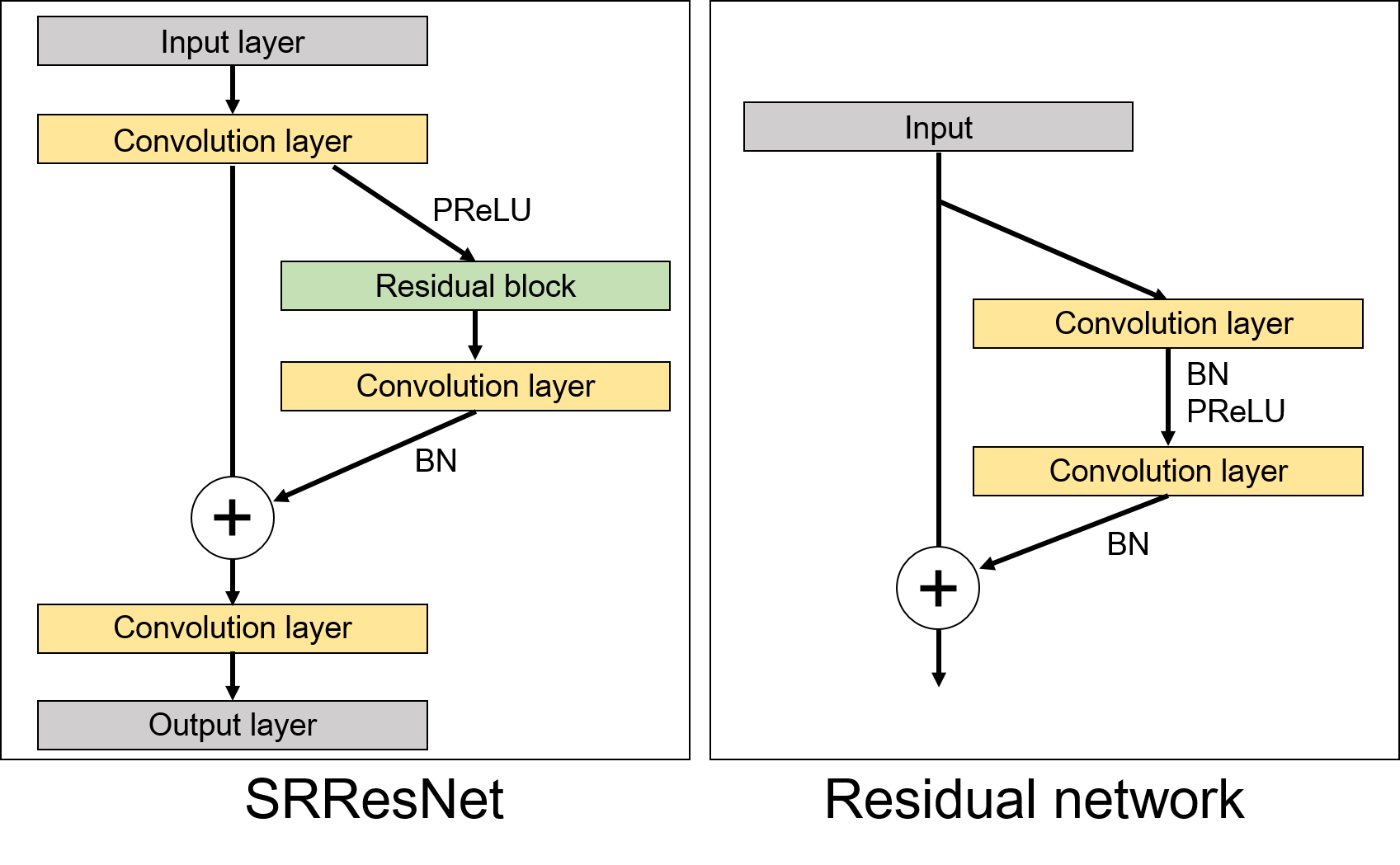}
\caption{\label{fig:resnet} ML  algorithms} architecture for Noise2Noise: the architecture of SRResNet (left) and residual block (right) are shown
\end{figure}
\begin{figure*}[htb]
\centering
\includegraphics[width = 1.0\linewidth]{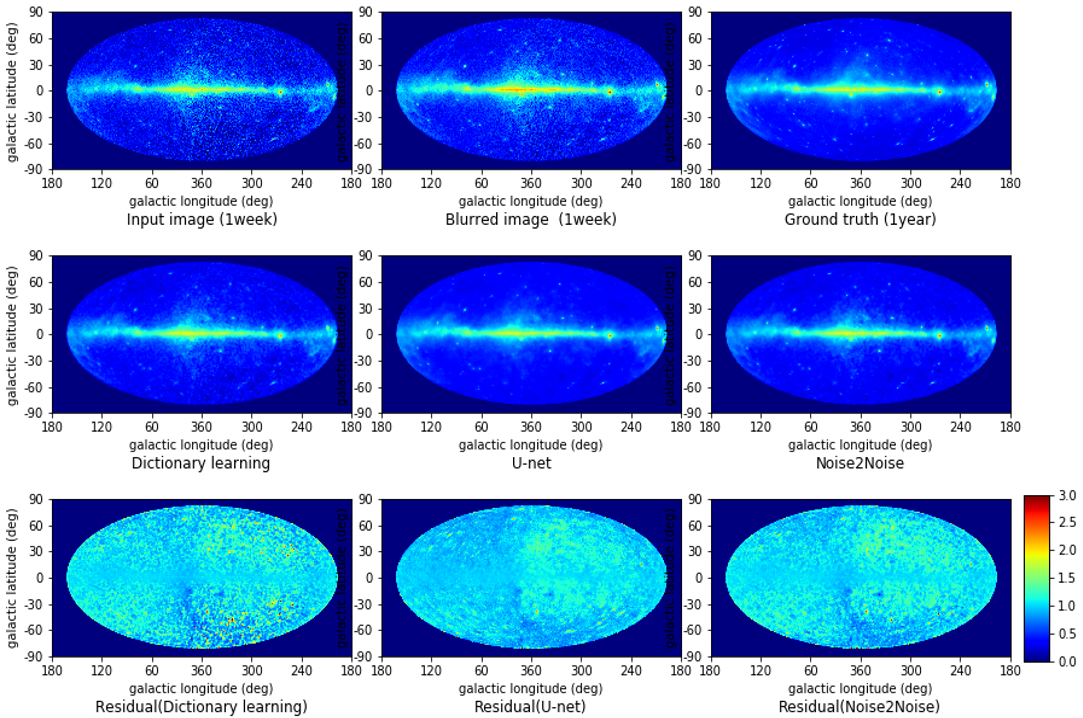}
\caption{\label{fig:all}  Results of applying ML algorithms to ASM obtained with Fermi-LAT. The input map, blurred map, and GT MAP are shown (top); ML maps generated by each ML algorithm (dictionary learning, U-net, and Noise2Noise) are shown (middle), and relative-ratio maps, which were calculated by dividing the GT MAP by each ML map are also shown (bottom).}
\end{figure*}

Hereinafter, “input map” and “GT map” refer to the ASMs, which cover the time interval MJD 58486.034–-58493.026 (2009 January 3--9) and MJD 58486.034–-58857.048 (2009 January 2–-2020 January 9), respectively. The input map was blurred with a 11 pixel $\times$ 11 pixel Gaussian kernel. In input map, standard deviation, which defines the smoothing effect, equals to 3.0. Additionally, we will use ”blurred map” to refer to blurred versions of the ASM, which were blurred with a 15 pixel $\times$ 15 pixel Gaussian blur kernel with a standard deviation equal to 100. “ML maps” refers to an ASM generated from the input map using three ML based algorithms (dictionary learning, U-net and Noise2Noise). We also note that the ML algorithm applied in this study only suggests the source variability in the corresponding image pixels. Therefore, detailed follow-up analysis such as drawing light curves of candidate sources was necessary to check the variability, as described in detail in Section~\ref{subsec:5-2}.
\begin{table}[ht]
\centering
\caption{\label{table1}Components of datasets prepared for training and testing of ML processing}
\begin{tabular}{llll}
\hline
                    &Training       &Validation     &Test\\
\hline
Sample number       &18000          &2000           &1\\
Observation year    &2009-2017      &2018           &2019\\
Extracted size(pixel)   &$64\times64$   &$64\times64$   &$3600\times1800$\\
Extracted size(degree)  &$6^{\circ}\times6^{\circ}$ &$6^{\circ}\times6^{\circ}$ &$360^{\circ}\times180^{\circ}$\\
\hline
\end{tabular}
\end{table}

\section{PREDICTION ALGORITHMS} \label{sec:3}
In this study, three  ML algorithms were used for prediction: dictionary learning, U-net, and Noise2Noise. The purpose and theory of these algorithms are detailed in the following sections.

\subsection{Dictionary learning}
Dictionary learning is generally used for dimensional reduction in the field of signal-processing \citep{Tosic_2011}. Unlike the deep-learning algorithms, dictionary learning is based on linear algebra; hence, one can sequentially trace the obtained results using this technique. Dictionary learning involves three steps: preparing a pair of databases, creating dictionaries, and generating images. First, five patches with 16 $\times$ 16 pixels were extracted randomly from each training sample of a short-observation map. The patches were arranged in columns and aligned to create a database of  short-observation data ($A_{\mathrm{short}}$). In the same manner as $A_{\mathrm{short}}$, a database of  long-observation data ($A_{\mathrm{long}}$) was also created. Then, the arrays, called the “dictionary,” were calculated. In this study, the dictionary from  short-observation data ($D_{\mathrm{short}}$) and  long-observation data ($D_{\mathrm{long}}$) were created from $A_{\mathrm{short}}$ and $A_{\mathrm{long}}$ by sparse representation. After the dictionaries were calculated, the test samples  from input map were represented by $D_{\mathrm{short}}$ and sparse matrix $C$. Finally, $D_{\mathrm{short}}$ was replaced with $D_{\mathrm{long}}$ to generate images corresponding to the GT map. The details of the calculation in dictionary learning are provided elsewhere \citep{SATO_2020_1}. In this study, scikit-learn (version 0.21.2), which is a library of  ML algorithms used in Python, was used for dictionary learning and sparse coding.

\subsection{U-net architecture} \label{subsec:3-2}
Originally, a U-net was developed for masking the medical images \citep{Han_2018, Dong_2017, Zhou_2018}. An example of the  algorithm architecture is illustrated in Figure~\ref{fig:unet}. U-net consists of two paths: the contracting path and expanding path. Two convolution layers and a max pooling layer are repeated four times in the contracting path. In this study, 3 $\times$ 3 convolution was used with leaky rectified linear unit (ReLU) activation and batch normalization. The number of channels was doubled for each downsampling. The expanding path consists of two convolution layers and a deconvolution layer with 3 $\times$ 3 filter sizes. The ReLU function was used as the activation function after the convolution layers except the last one. After the last convolutional layer, a hyperbolic tangent function was used. In this study, the "GeForce 1080Ti" graphics processing unit (GPU) processer was employed, and the U-net was implemented with TensorFlow (version 2.1.0) and Keras (version 2.3.1), which are libraries of deep-learning algorithms used in Python.

\subsection{Noise2Noise architecture}
Noise2Noise, which is a semisupervised learning  algorithms, is used to reconstruct clean signals from corrupted ones \citep{Dufan_2019, Hasan_2021, Baso_2019}. This  algorithms can perform signal denoising without the need for clean signal datasets. Therefore,  Noise2Noise is applicable to cases where clean signals are difficult to obtain. The superresolution residual network (SRResNet) \citep{Christian_2017} used in this study consists of a convolution layer with parametric rectified linear unit (PReLU) activation, six layers of residual block, two layers of convolution layer with batch normalization, and concatenated layer, as depicted in Figure~\ref{fig:resnet} (left). Figure~\ref{fig:resnet} (right) depicts the details of the residual block. The residual block consists of a convolution layer, batch normalization, and PReLU activation. In this study, the filter number and size in the convolution layer were set as 64 and 3 $\times$ 3, respectively. The software and GPU processer were used in the same way as for U-net in Section~\ref{subsec:3-2}.
\begin{figure}
\centering
\includegraphics[width = 1.0\linewidth]{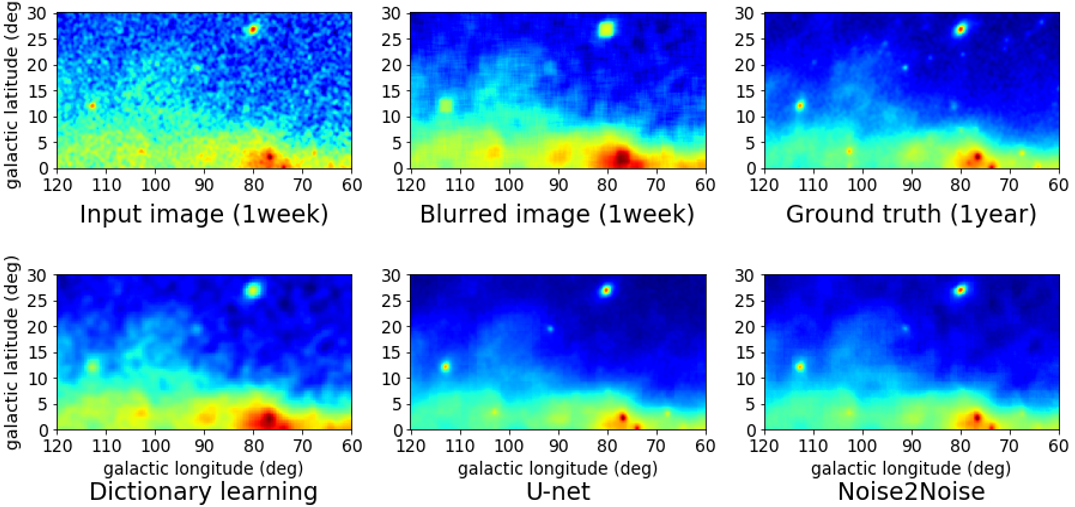}
\caption{\label{fig:ext1} Extracted part of input  map, blurred  map and GT map are shown (top), and the same parts of  ML maps are displayed (bottom)}
\end{figure}
\begin{figure}
\centering
\includegraphics[width = 0.9\linewidth]{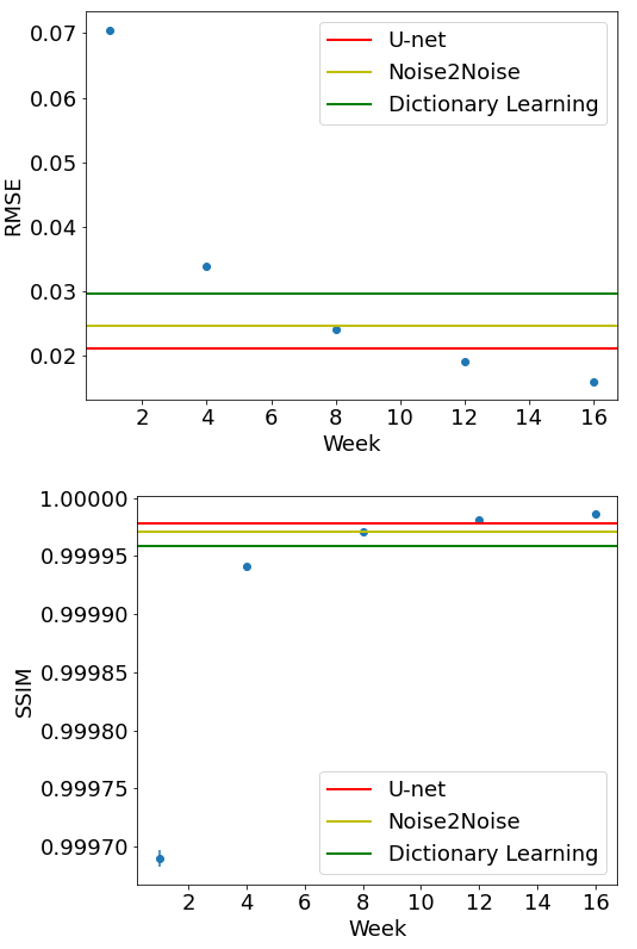}
\caption{\label{fig:eva}Evaluation results in RMSE (left) and SSIM (right) based on the GT map (one-year map). Maps with observation periods of one week, four weeks, eight weeks, 12 weeks, and 16 weeks were evaluated and plotted in blue. The result of applying ML maps are also shown.}
\end{figure}
\begin{figure*}
\centering
\includegraphics[width = 0.8\linewidth]{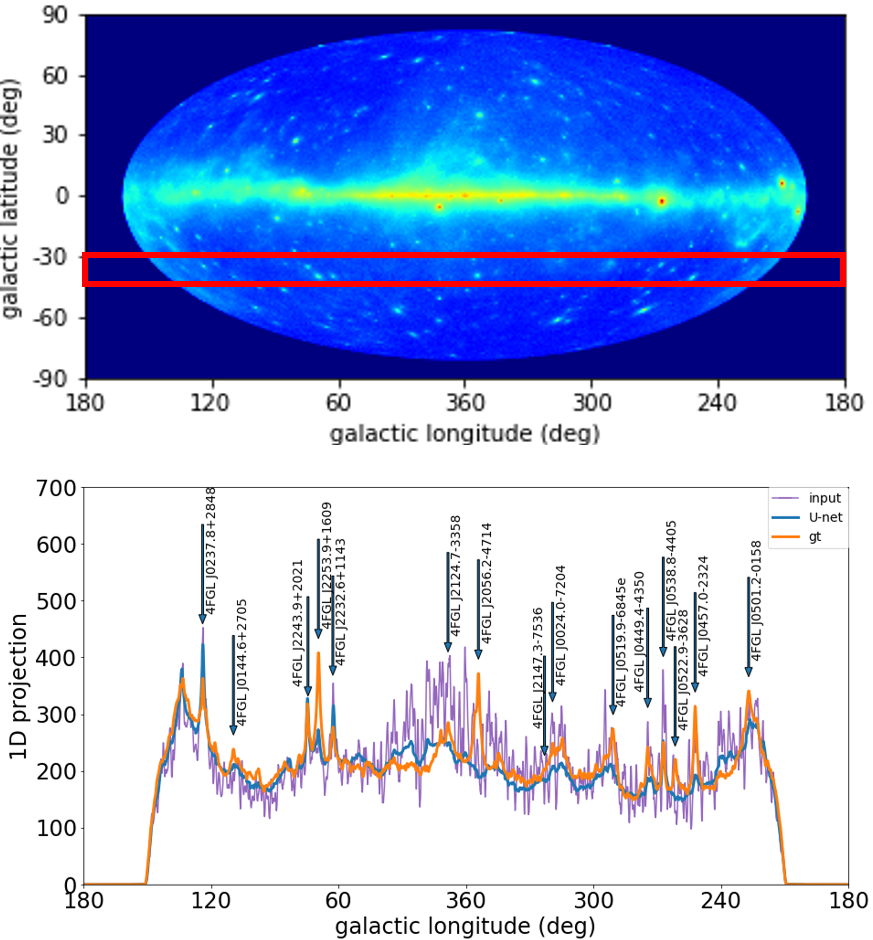}
\caption{\label{fig:4fgl}
Example of 1D projection histogram from input map, U-net map, and GT map. Extracted region (top) and 1D histogram (bottom) are shown. Representative sources in 4FGL catalog whose average flux are larger than $4\times10^{-9}$ are indicated.}
\end{figure*}

\section{Results}\label{sec:4}
\subsection{Improvement of image quality }\label{subsec:3-3}
To confirm the improvement of maps in detail, close-ups of the various regions in Figure~\ref{fig:all} were extracted, as shown in Figure~\ref{fig:ext1}. Image noise resulting from low photon statistics is reduced by ML algorithms; thus, the parts of the point source are enhanced. Moreover, feature of the diffuse gas emission around the galactic plane is also well reproduced.

To evaluate the image improvement in ML maps, we applied a root mean squared error (RMSE), which is the standard evaluation index in computer vision. The RMSE represents the squared error of each pixel value in the two images and was calculated using the following equation:
\begin{equation}
\label{eq:3-1}
\mathrm{RMSE}=\sqrt{\frac{1}{n}\Sigma_{\mathrm{i}=1}^{\mathrm{n}}(V_{\mathrm{i}}^{\mathrm{image1}}-V_{\mathrm{i}}^{\mathrm{image2}})^{2}},
\end{equation}
where $\mathrm{n}$ and $V_{\mathrm{i}}$ denote the number of image pixels and the $i$-th pixel value in each input image, respectively. An RMSE that is closer to zero represents a smaller error between the input images. Similarly, we also computed the structural similarity (SSIM) between the produced maps and the GT map. The SSIM, which represents visual similarity, was calculated using equation~\ref{eq:3-2}.
\begin{equation}
\label{eq:3-2}
\mathrm{SSIM}=\frac{(2\mu_{x}\mu_{y}+\mathrm{C}_{1})(2\sigma_{xy}+\mathrm{C}_{2})}
{(\mu_{x}^{2}+\mu_{y}^{2}+\mathrm{C}_{1})(\sigma_{x}^{2}+\sigma_{y}^{2}+\mathrm{C}_{2}),},
\end{equation}
where $\mu$ and $\sigma$ denote the average value and standard deviation of the image, respectively. $\mathrm{C}_{1}$ and $\mathrm{C}_{2}$ are small constants for avoiding instability in case that $\mu_{x}^{2}+\mu_{y}^{2}$ or $\sigma_{x}^{2}+\sigma_{y}^{2}$ is close to zero. Details of the constants are described in Wang et al.\citep{Wang_2004}. Figure~\ref{fig:eva} depicted the results of RMSE and SSIM evaluation. In Figure~\ref{fig:eva}, the horizontal axis represents the observation time, and the vertical axis represents the RMSE or SSIM values. We discussed the evaluation results in Section~\ref{subsec:5-3}.
\begin{figure}
\centering
\includegraphics[width = 1.0\linewidth]{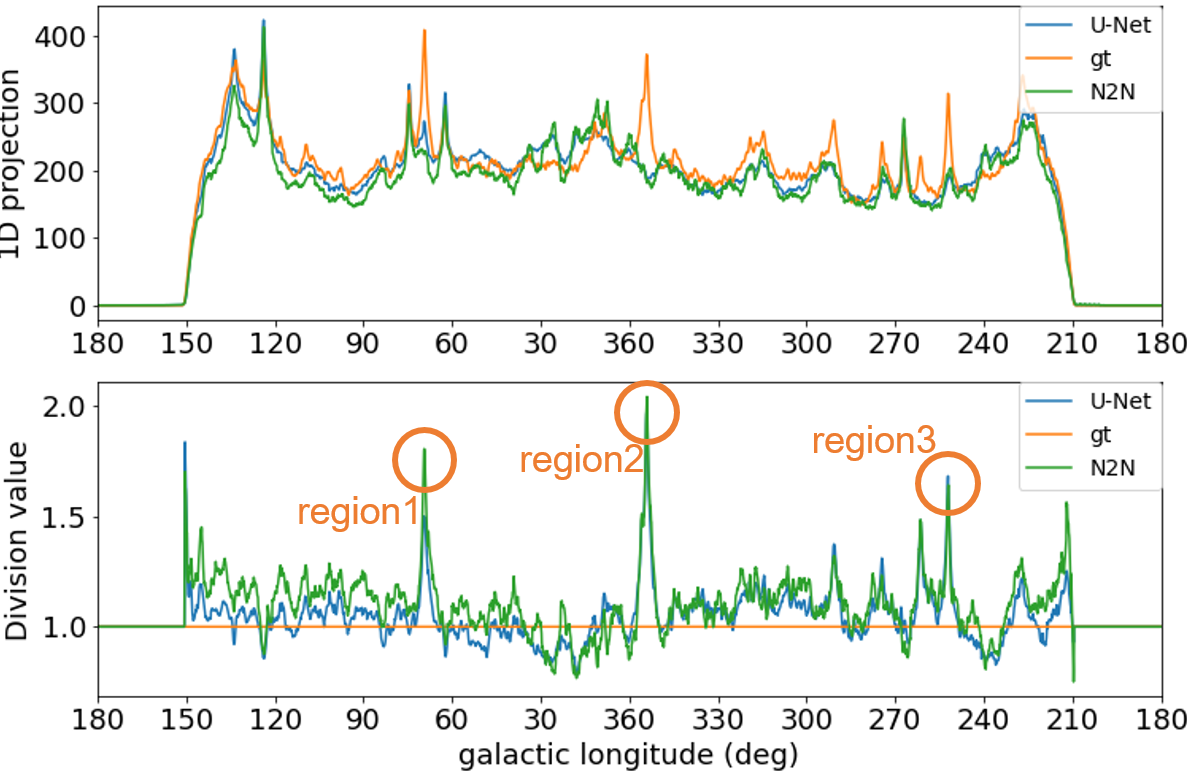}
\caption{\label{fig:proj1}  Example of projection histogram and division: 1D projection (top) and the division values obtained by dividing GT map by the values of ML maps are shown (bottom), respectively. Orange circles show the large difference between the 1D projection calculated from the ML maps and the projection from GT map}
\end{figure}
\begin{figure}
\centering
\includegraphics[width = 1.0\linewidth]{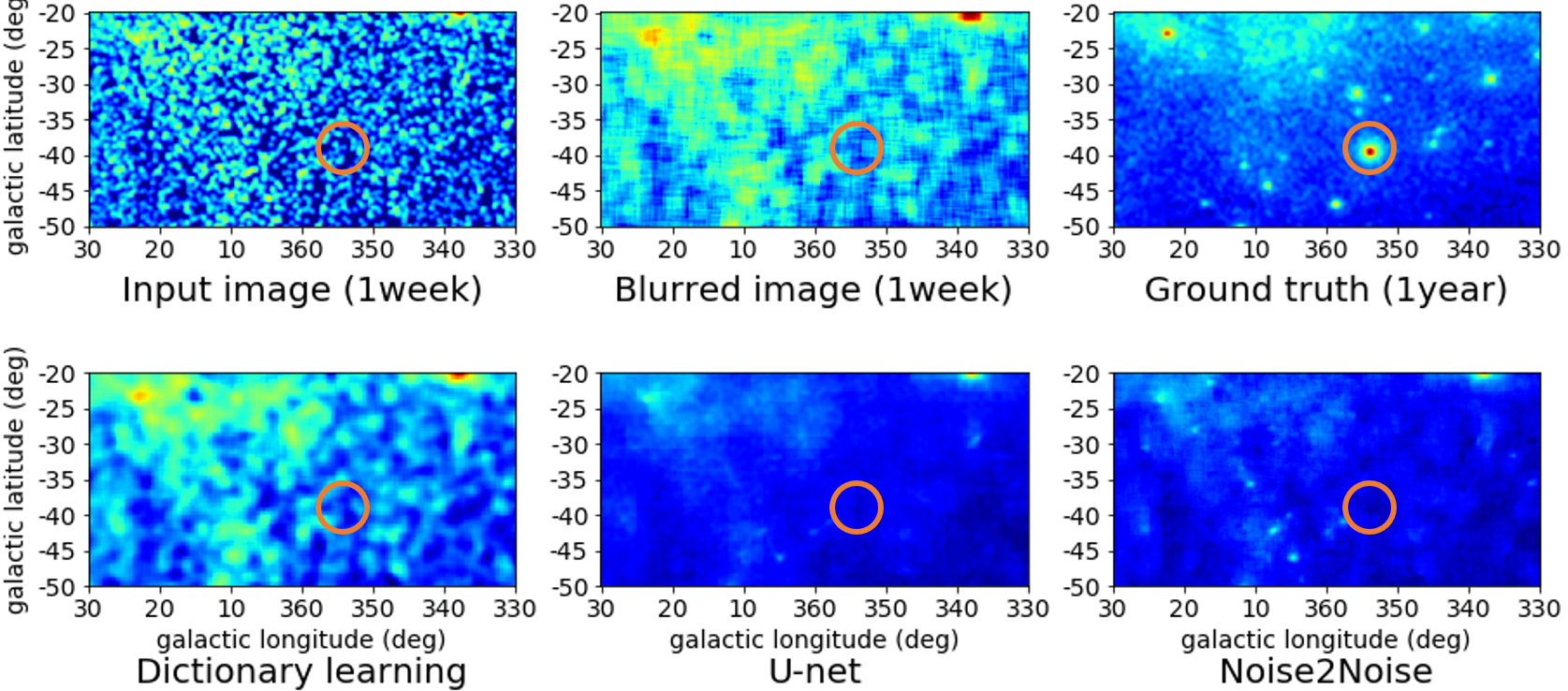}
\caption{\label{fig:ext2} Corresponding part of large division value (PKS2052-47): input  map, blurred  map, and GT map are shown (top), the same parts of the ML maps are displayed (bottom), and the part with a large difference is surrounded by an orange circle}
\end{figure}

\subsection{Comparing ML maps to GT map}
\begin{figure}
\centering
\includegraphics[width = 1.0\linewidth]{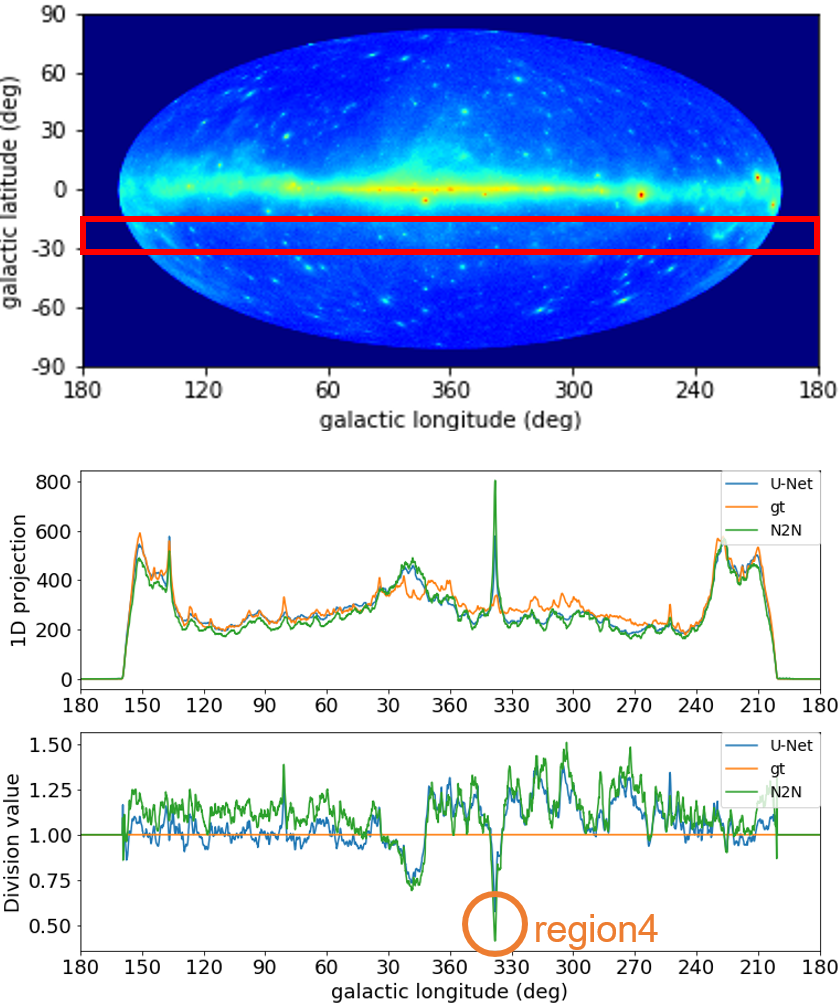}
\caption{\label{fig:proj2} Negative case of projection histogram and division: the extracted part (top) and the 1D projection (middle) are shown, the division values obtained by dividing GT map by  the values of ML maps are shown (bottom), and orange circles show the large difference between the 1D projection calculated from  ML maps and the projection from GT map}
\end{figure}
\begin{figure}
\centering
\includegraphics[width = 1.0\linewidth]{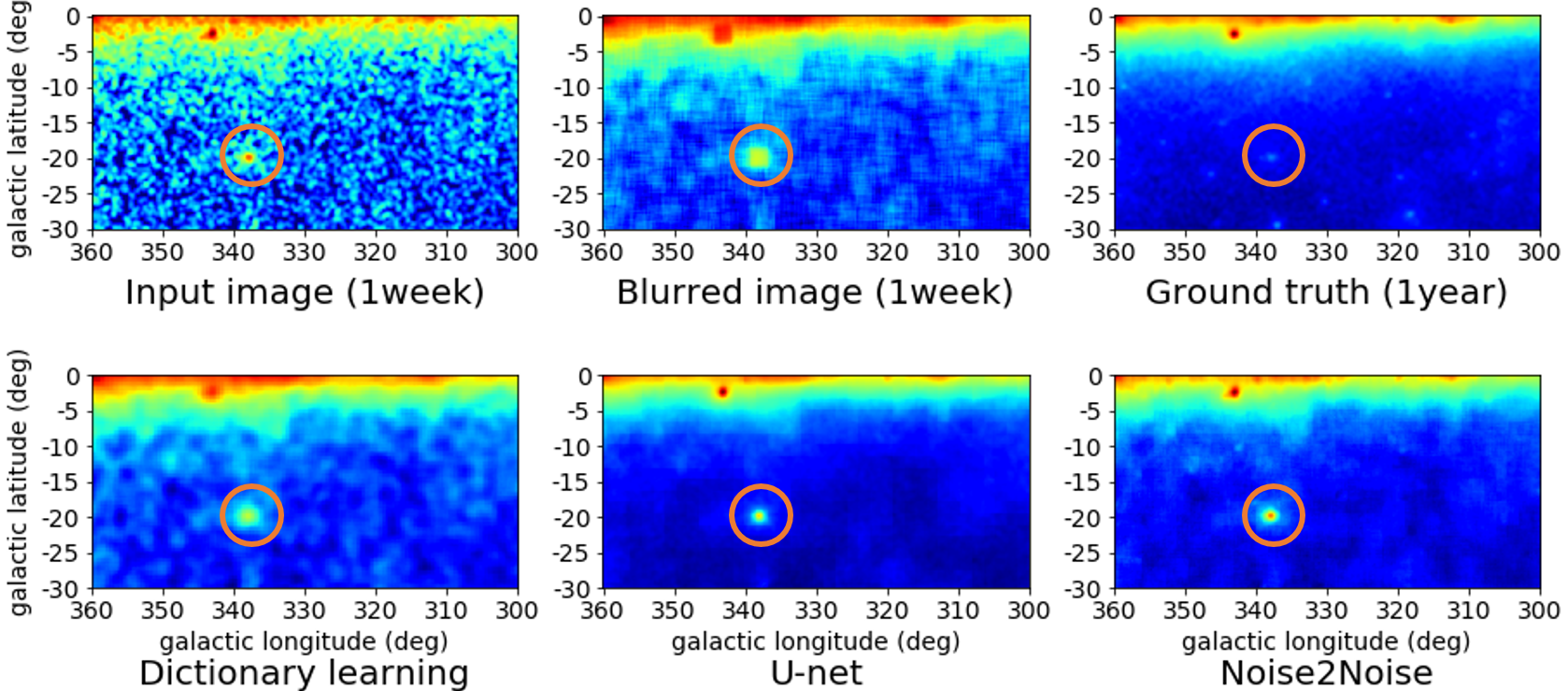}
\caption{\label{fig:ext3} Corresponding part of large division value (PKS1824-582): input  map, blurred  map, and GT map are shown (top), the same parts of the  ML maps are displayed (bottom), and the part with a large difference is surrounded by an orange circle}
\end{figure}
\begin{table*}[ht]
\centering
\caption{\label{table2}List of source IDs and positions corresponding to the circled part in Figures~\ref{fig:proj1} and~\ref{fig:proj2}}
\begin{tabular}{lllllll}
\hline
region  &assoc4FGL  &source name    &source type&redshift   &RA     &Dec\\
        &           &               &           &           &J2000  &J2000\\
\hline
region1 &3C454.3    &J2253.9$+$1609 &AGN/FSRQ   & 0.859 &343.4906   &16.1482\\
region2 &PKS2052-47 &J2056.2$-$4714 &AGN/FSRQ   & 1.489 &314.0715   &$-$47.2369\\
region3 &PKS0454-234&J0522.9$-$3628 &AGN/FSRQ   & 1.003 &74.2608    &$-$23.4145\\
region4 &PKS1824-582&J1829.2$-$5813 &AGN/FSRQ   & 1.531 &277.3108   &$-$58.2323\\
\hline
\end{tabular}
\end{table*}
For source detection, we compared the excess of pixel values in the ML maps corresponding to the source position, as already listed/known in the 4FGL catalog. Figure~\ref{fig:4fgl} (bottom) shows an example of a projection histogram, the region of which is shown in top panel. In Figure~\ref{fig:4fgl}, 15 representative sources with the highest average flux are indicated. The signal-to-noise ratio (S/N) in the projection was improved by ML algorithms. However, unlike input and GT maps based on photon statistics, pixel values in the ML maps lose any physical meaning; thus, we could not evaluate test statistics (TS), which is often applied in the analysis of Fermi LAT to evaluate the significance of source detection. We will develop such advanced algorithms for ML maps in the future; however, it is beyond the scope of this paper.

To find the correspondence between the  ML maps and GT map, the 1D projections along the same galactic latitude $l$ were compared for various regions. An example of a projection histogram obtained by GT map, U-net map, and Noise2Noise map is depicted in Figure~\ref{fig:proj1} (top). The extracted part is the same as in Figure~\ref{fig:4fgl}. The projection was created by integrating the pixel values for 15$^{\circ}$ in the direction of galactic latitude $l$. For a quantitative comparison, the values in the ML maps were divided by those in the GT map. Figure~\ref{fig:proj1} (bottom) indicates the ratio of the ML maps to the GT map. Three regions with large differences are marked by orange circles. Figure~\ref{fig:ext2} shows close-ups around region 2 to illustrate such a difference. While no evident sources exist in the input map and  ML maps, a bright point source, which is identified as PKS2052-47, is found only in the GT map. In the opposite case shown in Figure~\ref{fig:proj1}, there exist regions in which the ratio is smaller than unity, as represented by region 4 in Figure~\ref{fig:proj2}. Figure~\ref{fig:ext3} shows close-ups around region 4. The details of regions with apparent ”mismatch” are listed in Table~\ref{table2}.

\section{DISCUSSION} \label{sec:5}
\subsection{Comparing various ML algorithms} \label{subsec:5-1}
In this study, three types of  ML algorithms were applied: dictionary learning, U-net, and Noise2Noise. Dictionary learning is based on linear algebra and can be used to trace the calculation process sequentially. However, dictionary learning has a low representation capability compared with other deep-learning algorithms and reflects a relatively low-quality image compared with other algorithms presented in Figure~\ref{fig:ext1}. Thus, dictionary learning is less effective than other deep-learning algorithms, for the purpose of this study. In the U-net architecture, the input and output layers are connected by two paths: an extended path and a shortcut connection. Since these two paths are considered to capture detailed features and context, U-net can successfully predict  long-observation map from  the input map with high precision (see Figures~\ref{fig:ext1}). However, in Noise2Noise, the  algorithm captures features of statistical noise by comparing two types of short-observation maps. Therefore, the  algorithm is used to generate the ASM by neglecting the statistical noise, which can sometimes overestimate the significance of each point source. Thus, U-net may be superior to Noise2Noise for predicting a  long-observation map. In this study, U-net was considered to be the most reliable/precise for predicting the  long-observation map. Furthermore because these three  ML algorithms are based on completely different theories,  ML maps are more reliable if the outputs from each  ML algorithm are similar to each other.

\begin{figure*}[htb]
\centering
\includegraphics[width = 1.0\linewidth]{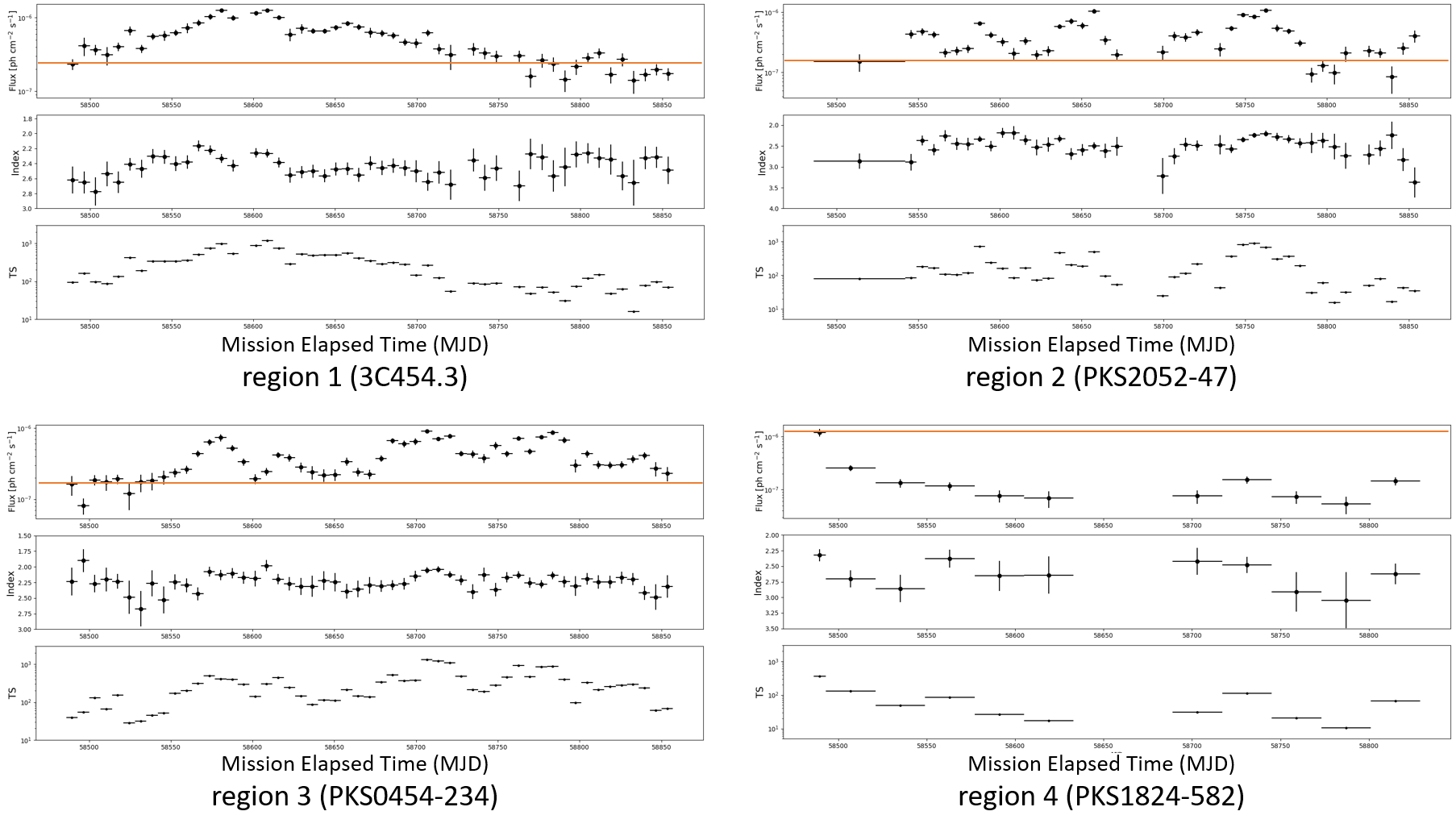}
\caption{\label{fig:lc}
Fermi-LAT light curve and variation of photon index for each source shown in Table~\ref{table2} over the time interval from January 2019 to December 2019 in one-week time bins. In each graph, only flux bins with test statistic TS $>$ 9 is plotted. Because PKS2052-47 had too few photons and the TS did not exceed 9, the data for the first two months from the beginning of 2019 were used as the first bin. Similarly, the light curve for PKS1824-582 has few bins with TS $>$ 9 throughout the year, so its light curve data were binned with one-month intervals, while only the first bin was binned with one-week data because the number of photons was sufficient for the first week.  Top panel: changes in the $100$MeV $< E < 300$ GeV flux. The orange line shows the initial one-week flux.  Middle panel: changes in the power-law photon index.  Bottom panel: changes in the TS.
}
\end{figure*}

\subsection{Origin of mismatch between GT map and  ML maps} \label{subsec:5-2}
In this section, a possible reason for the mismatch between the GT map and ML maps, as shown in Figures ~\ref{fig:proj1} and~\ref{fig:proj2}, is discussed. First, the regions of the largest mismatch, indicated as orange circles in Figures~\ref{fig:proj1} and~\ref{fig:proj2}, are discussed. The ML maps are generated based on the hypothesis that the sources are persistent over one year, which is, however, not the case for variable and/or transient sources. Thus, the observed mismatch between the  ML maps and GT map is likely associated with the variable and transient sources, such as the flaring activity of AGNs. To confirm this assumption, the sources listed in Table~\ref{table2} are analyzed in more detail with the Fermi analysis tool, as described in section~\ref{subsec:2-1}. The light curves around each source are shown in Figure~\ref{fig:lc}. For example, a variable source in region 1 corresponds to a flat spectrum radio quasar (FSRQ) 3C454.3 with a redshift of $z$ = 0.859 (Table~\ref{table2}). As shown in Figure~\ref{fig:lc}, the flux of 3C454.3 in the first week was much smaller than the average flux, shown as an orange line, integrated over a year. Similarly, regions 2 and 3 were identified as two FSRQs; PKS2052-47 and PKS0454-234, respectively. Contrary to the sources in which flux substantially increased after the first week, the average flux over the year is smaller than the first-week’s flux in the case of region 4, which corresponds to FSRQ PKS1824-582. Obviously, such temporal variability in fluxes explains the observed mismatch between the ML maps and GT map, as shown in Figures ~\ref{fig:proj1} and~\ref{fig:proj2}. As can be seen in this figure’, point sources with increasing photon flux appear observed to be brighter in Figure~\ref{fig:all} (bottom). On the contrary, point sources with decreasing flux appear faint.

However, there is a systematic difference between the GT map and  ML maps around the galactic longitude 270$^{\circ}$  $<$ $l$ $<$ 330$^{\circ}$ in Figure~\ref{fig:proj1} (bottom), which cannot be explained either by the statistical fluctuation or combination of variable sources.  We also confirm the difference in Figure~\ref{fig:all} (bottom). Such a systematic gradient is thought to be caused by the nonuniformity of the exposures inherent in the one-week observation, which were used as input for  ML algorithms. Although exposure map correction is performed based on the position and angle of the satellite, such nonuniformity affects the reconstructed counting map, especially in the case of short observation times. In fact, we confirmed that such a gradient is reduced if two-week ASM maps are used as input, but the optimization of the time duration for the input maps is beyond the scope of this study.  Additionally, ASMs in this paper were produced with no energy binning, hence the exposure map might not correct the actual exposure. This is because the accuracy of integrating the exposure over the energy band depends on the smoothness of the source spectrum across the energy band, and these are quite large energy bands.

\subsection{Quality of ML maps} \label{subsec:5-3}
As shown in Figure~\ref{fig:eva}, the ML-based algorithms were effective in improving the ASM quality. In particular, U-net was the best among the three ML algorithms. The RMSE and SSIM assessment of the U-net map were equivalent to approximately 10 and 12 weeks, respectively. Considering that the exposure-map and variable source were incorrect, the image quality was improved by more than 10 weeks. Despite the significant improvement, faint source detection was considered difficult for our algorithms due to image blur. Some peaks around the source location were blurred, as shown in Figure~\ref{fig:4fgl}. In general, for ML processing, the loss function must be defined. “Training ML algorithms” refer to parameter optimization, which could be used for minimizing the loss function. The RMSE, which is one of the standard loss functions, was used in this study. However, in the case of using MSE or RMSE, it is known that high-frequency components are difficult to reproduce. Due to the difficulty of point-source detection, some schemes are necessary to improve the detection efficiency. For example, adopting the sum of point source detectability and RMSE as the loss function may improve the performance. Other approaches are mentioned in Section~\ref{subsec:5-5}.

\subsection{Comparison with existing method} \label{subsec:5-4}
Compared with other techniques described in section\ref{sec:intro}, our ML-based method has the following three advantages: 
\begin{enumerate}
    \item The ML algorithm can be easily implemented by end users. Moreover, any assumption or modeling of foreground diffuse emission is unnecessary. Owing to its simplicity, the ML all-sky map can be obtained within less than 10 min.
    \item Signal-to-noise on the image is improved both in terms of RMSE and SSIM, although the optimum algorithms to detect faint point sources remain a future work.
    \item By comparing the ML and GT maps, variable sources can be detected automatically in ASM in either cases (increasing and/or decreasing) of fluxes.
\end{enumerate}
Conversely, the ML method generally loses the physical properties of the source images, such as the energy of each photon and the statistical significance. Thus, one should consider such an ML-based image as a starting point for finding possible interesting sources. These sources would be chosen for subsequent detailed analysis based on the standard analysis tool released by Fermi-LAT collaboration. Detailed modeling of DGE, as well as evaluation of the test statistic (TS), is essential even after the application of ML methods presented in this paper.

In this context, the D3PO inference algorithm, which is one of the latest modeling techniques, demonstrated excellent noise reduction. However, analysis with complex models is required, which is labor-intensive and computationally expensive. Our ML-based method, on the contrary, is model-independent, and can be completed by simple implementation, as mentioned above. While the method is cost-effective, ML maps lose physical properties, such as energy spectrum.

Finally, we discussed the detection variable source in comparison to existing methods. For source variability in the 4FGL catalog, the variability index, such as $\mathrm{TS}_{\mathrm{var}}$ \citep{Nolan_2012}, was calculated from light curves using one-year bin data. For bright and variable sources, light curves are automatically updated daily for cataloged or well-known sources. In contrast, our ML methods have the potential to automatically detect serendipitous variable sources. However, as mentioned earlier, our ML-based technique loses any physical properties of the sources; hence, a detailed analysis is necessary. In this context, FAVA, which is a simple and effective method for detecting variable sources, was employed. However, as FAVA detects the deviation between the map expected from long-term observations and the one from short-term observations, the S/N does not change. Our algorithms improved the S/N around the point source; hence, the variable source detection has the potential to achieve higher accuracy than FAVA.

\subsection{Future prospects} \label{subsec:5-5}
In this study, an initial one-week map obtained with Fermi-LAT in 2019 was prepared as a test sample. However, the prediction method is applicable not only to gamma-ray sky maps but also to other maps obtained in different energy bands, such as radio, optical, and X-ray ASMs. Moreover, one can set any time interval and duration depending on the research purpose, astrophysical phenomenon, and source. For example, one can use a time scale that is as short as minutes, to detect various transient sources, such as GRBs, without human bias at any point by applying ML algorithms. Longer time scales of weeks to months are more favorable for searching tidal disruption events \citep{Burrows_2011, Gezari_2012} and/or supernova.

In this research, a one-week map was used as input map to demonstrate the concept and performance of various  ML algorithms as applied to Fermi-LAT data. In planned future work, an attempt will be made to find transient sources by setting  ASMs generated at different time scales, to widely cover scales of minutes to days as input. Moreover, another interesting possibility is that one can predict longer observation maps if the observations are continued, even after the mission has been completed. For example, if a one-year ASM is used as the input, it may be possible to predict a 100-year ASM from existing data. Clearly, such predictions may or may not be correct, as shown in Section~\ref{sec:5}; however, it can provide strong motivation for future missions and important scientific aspects to be explored in the future. Similar ML approaches have been proposed to confirm the giant but weak emission structure of diffuse gas, such as the Fermi bubble \citep{Su_2010} and NPS/Loop I \citep{Kataoka_2018}, which was proposed using a Gaussian mixture  algorithms \citep{Muller_2018}.

For future research directions, three different approaches are promising. The first is to train the ML algorithm separately for the point source and DGE regions. In this study, although we used ASM data for training, the DGE and point source regions were separated to train the ML algorithms. The second approach is to apply other  ML algorithms, especially deep-learning algorithms. In this study, U-net and Noise2Noise were considered to be effective for improving the ASM obtained by Fermi-LAT. However, it is necessary to search other  ML algorithms, such as the generative adversarial network (GAN) type or Fourier convolutional neural network (FCNN)-type for more practical applications. By using the FCNN-type  algorithms, it may be possible to obtain the features represented in the frequency domain. Another approach is to use a large amount of data for training. In this study, only actual observational data were prepared. The amount of training data can be easily increased using simulation data.

\section{CONCLUSION} \label{sec:6}
A novel application of ML algorithms to analyze ASMs obtained with the Fermi-LAT was demonstrated in this paper. An ASM corresponding to a long- observation map was successfully generated from short- observation map by using three types of ML algorithms. ML based methods make it possible to improve image quality, not only for point sources, but also for diffuse gas structures. Furthermore, these methods are useful for detecting variable sources automatically without human bias by analyzing large deviations between the simulated ML map and GT map. In addition, the method can be extended to other satellite data, even if the satellite has completed its observations. Thus, in future work, the proposed method will be applied to other satellite data, such as eROSITA on SRG and/or BATSE onboard CGRO.

\acknowledgements

 We thank an anonymous referee for his/her careful advice to improve the manuscript. This research was supported by Japan Society for the Promotion of Science (JSPS) KAKENHI [Grant Number 20H00669].

\bibliography{bibs}{}
\bibliographystyle{aasjournal}
\end{document}